%% file: main.tex
\pdfoutput=1
\documentclass[11pt]{article}

\usepackage[final]{acl}

\usepackage{times}
\usepackage{latexsym}

\usepackage[T1]{fontenc}
\usepackage[utf8]{inputenc}

\usepackage{microtype}

\usepackage{inconsolata}

\usepackage{graphicx}
\usepackage[textsize=tiny,textwidth=2.3cm,disable]{todonotes}
\usepackage{booktabs}
\usepackage{enumitem}
\title{TrustMap: Mapping Truthfulness Stance of Social Media Posts on Factual Claims for Geographical Analysis}

\author{Zhengyuan Zhu, Haiqi Zhang, Zeyu Zhang, Chengkai Li \\
University of Texas at Arlington \\
\texttt{\{zhengyuan.zhu, haiqi.zhang, zeyu.zhang\}@mavs.uta.edu}\\ \texttt{cli@uta.edu}}

\begin{document}
\maketitle
\input{macro}

\input{sec-abstract}
\input{sec_intro}
\input{sec-related}
\input{sec-design}
\input{sec-result}

\input{sec-conclusion}

\bibliography{custom}

\input{sec-appendix}

\end{document}

%% file: macro.tex
\newcommand{\system}[1]{{\small \ensuremath {\mathsf{#1}}}}
\newcommand{\TSDCT}{\system{TSD}-\system{CT}}
\newcommand{\RATSD}{\system{RATSD}}
\newcommand{\TRUSTMAP}{\system{TrustMap}}
\newcommand{\highlight}[1]{\colorbox{yellow}{#1}}

%% file: sec-abstract.tex
\begin{abstract}
Factual claims and misinformation circulate widely on social media and affect how people form opinions and make decisions. This paper presents a \textbf{tru}thfulness \textbf{st}ance \textbf{map} (\TRUSTMAP), an application that identifies and maps public stances toward factual claims across U.S. regions. Each social media post is classified as positive, negative, or neutral/no stance, based on whether it believes a factual claim is true or false, expresses uncertainty about the truthfulness, or does not explicitly take a position on the claim’s truthfulness. The tool uses a retrieval-augmented model with fine-tuned language models for automatic stance classification. The stance classification results and social media posts are grouped by location to show how stance patterns vary geographically. \TRUSTMAP\ allows users to explore these patterns by claim and region and connects stance detection with geographical analysis to better understand public engagement with factual claims. 
\end{abstract}

%% file: sec_intro.tex
\section{Introduction}

In the digital era, the dissemination of factual claims, public narratives, and misinformation has intensified across various domains, including health~\cite{suarez2021prevalence}, environment~\cite{treen2020online}, and politics~\cite{tucker2018social}. Social media platforms such as X (formerly Twitter) play an important role in online discourse by enabling the broad spread of content~\cite{bakshy2012role, shi2014content}. 
However, these platforms also facilitate the proliferation of misleading claims and conspiracy theories, which can significantly influence public opinions and decision-making processes~\cite{allcott2019trends, ausat2023role, yan2025origin}. In addition, these online discussions not only reflect public sentiment but also affect perceptions of the truthfulness of factual claims, critically shaping awareness and engagement with key societal issues. 

\begin{figure}[t]
    \begin{center}    	
    \includegraphics[width=0.91\columnwidth]{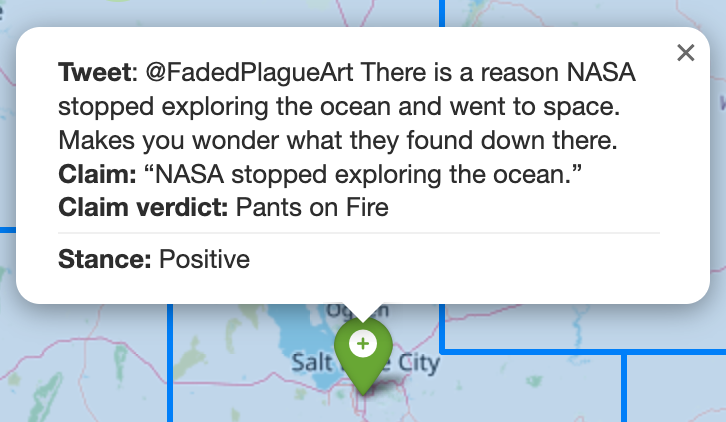}
    \caption{An example of a claim-tweet pair and its truthfulness stance displayed on \TRUSTMAP.}
    \label{fig:stance_example}
    \end{center}
\end{figure}

Social media users frequently respond to factual claims by endorsing their veracity, disputing their validity, or expressing uncertainty. 
Understanding these responses, or truthfulness stances, is essential for analyzing how misinformation and public messaging shape discussions across different topics. 
While previous research on stance detection has explored misinformation in health~\cite{hossain2020covidlies}, news~\cite{reddy2021newsclaims}, and other controversial topics~\cite{zhang-etal-2024-granular}, many studies have not incorporated geolocation-based stance analysis across diverse subject areas. The ability to analyze stance perceptions at regional and local levels is crucial for assessing the influence of narratives on different communities.

This paper introduces the \textbf{tru}thfulness \textbf{st}ance \textbf{map} (\TRUSTMAP), a system that visually presents stance distributions toward factual claims across geographical regions in the United States. It aggregates social media users’ perceptions of truthfulness and classifies them into three stance categories---positive, negative, and neutral/no stance, based on the conceptual framework from~\citet{ratsd-naacl25}.
A positive stance indicates that a post conveys support for a claim’s veracity, a negative stance suggests that the post disputes the claim as false, and a neutral/no stance signifies either uncertainty or a lack of explicit endorsement or refutation. 
Figure~\ref{fig:stance_example} shows an example of a tweet presenting a positive truthfulness stance toward a claim displayed on \TRUSTMAP. 

We collected 136,040 tweets related to 2,216 distinct claims. Among these, 24,262 tweets include geolocation information. Each claim-tweet pair is classified using \RATSD\ (Retrieval
Augmented Truthfulness Stance Detection)~\cite{ratsd-naacl25}. This model leverages fine-tuned large language models (LLMs) and retrieval-augmented generation (RAG)~\cite{lewis2020retrieval}. The truthfulness stance detection results are rendered on \TRUSTMAP. In \TRUSTMAP, users can select one or multiple topics, choose factual claims within the selected topic(s), explore tweets related to the selected claims based on their truthfulness stance with respect to the selected claims, and visualize stance distributions across geographical areas and over time. 

The result analysis suggests that social media users often believe claims are true regardless of their actual veracity. From a topical perspective, users show the poorest judgment on claims related to the environment. Geographically, users in Florida demonstrate the greatest difficulty in distinguishing between true and false claims.

To the best of our knowledge, \TRUSTMAP\ is the first publicly available map application that visualizes truthfulness stances across different topics. In summary, this paper's contributions are as follows: 
\begin{itemize}[noitemsep,wide,topsep=0pt]
    \item We developed a truthfulness stance map (\href{https://trustmap.streamlit.app}{https://trustmap.streamlit.app}) for visualizing stance distributions across various topics and regions at varying levels of granularity (i.e., states and cities). A video demo of \TRUSTMAP\ is available at \href{https://vimeo.com/1070530797}{https://vimeo.com/1070530797}.
    \item We evaluated the truthfulness stance detection model and analyzed stance classification results across different topics and regions.
    \item The codebase and data are available at \href{https://github.com/idirlab/trustmap}{https://github.com/idirlab/trustmap} to promote research reproducibility and facilitate further studies on geographical truthfulness stance analysis.
\end{itemize}

% \begin{figure*}[h]
%     \begin{center}
%     	\includegraphics[width=1.6\columnwidth]{Figures/stance_map_diagram.png}
% 		\caption{The system architecture of truthfulness stance map.}
% 		\label{fig:diagram}
%     \end{center}
% \end{figure*}

%% file: sec-related.tex
\section{Related Work}

Although many existing systems focus on online narrative monitoring and analytics, most of them emphasize general data collection and analytics~\cite{borra2014programmed,zhang2024wildfire} or popular domains such as sentiment analysis~\cite{lien2022osn,agarwal2018geospatial}. The incorporation of truthfulness stance into interactive maps or dashboards remains a relatively new concept. Existing recent studies primarily explore stance types different from those examined in our research.  StanceVis Prime~\cite{kucher2020stancevis} is designed for the analysis and visualization of sentiment and stance in temporal text data from various social media sources. It processes documents from multiple text streams and applies sentiment and stance classification, generating data series linked to the source texts. However, its stance detection is based on identifying seven different modifiers~\cite{skeppstedt2017detection} in a given text, which differs from our definition of truthfulness stance. \citet{liew2024designing} designed a discourse analysis dashboard for monitoring and analyzing online narratives. This dashboard focuses on \emph{sentiment}-oriented stance toward \emph{general} topics such as vaccine side effects, which is also distinct from our focus on \emph{truthfulness}-oriented stance toward \emph{individual} claims. 

Existing systems focusing on truthfulness stance detection primarily stem from our previous work. We developed a dashboard for the COVID-19 misinfodemic~\cite{zhu2021dashboard}, which identified stances toward COVID-19 related facts. However, stance detection was not the primary focus of that system, and the underlying method was less advanced than \RATSD. In contrast, \TRUSTMAP\ supports more sophisticated stance detection and covers a broader range of factual claims, including both verified facts and misinformation across multiple topics beyond COVID-19.
We also built a framework to understand social media users' truthfulness stance toward claims across climate change-related topics~\cite{zhang-etal-2024-granular}, and we developed novel truthfulness stance detection methodologies~\cite{ratsd-naacl25}.
These previous works contribute to the foundation of \TRUSTMAP. 

%% file: sec-design.tex
\begin{figure*}[t]
    \begin{center}
    	\includegraphics[width=1.97\columnwidth]{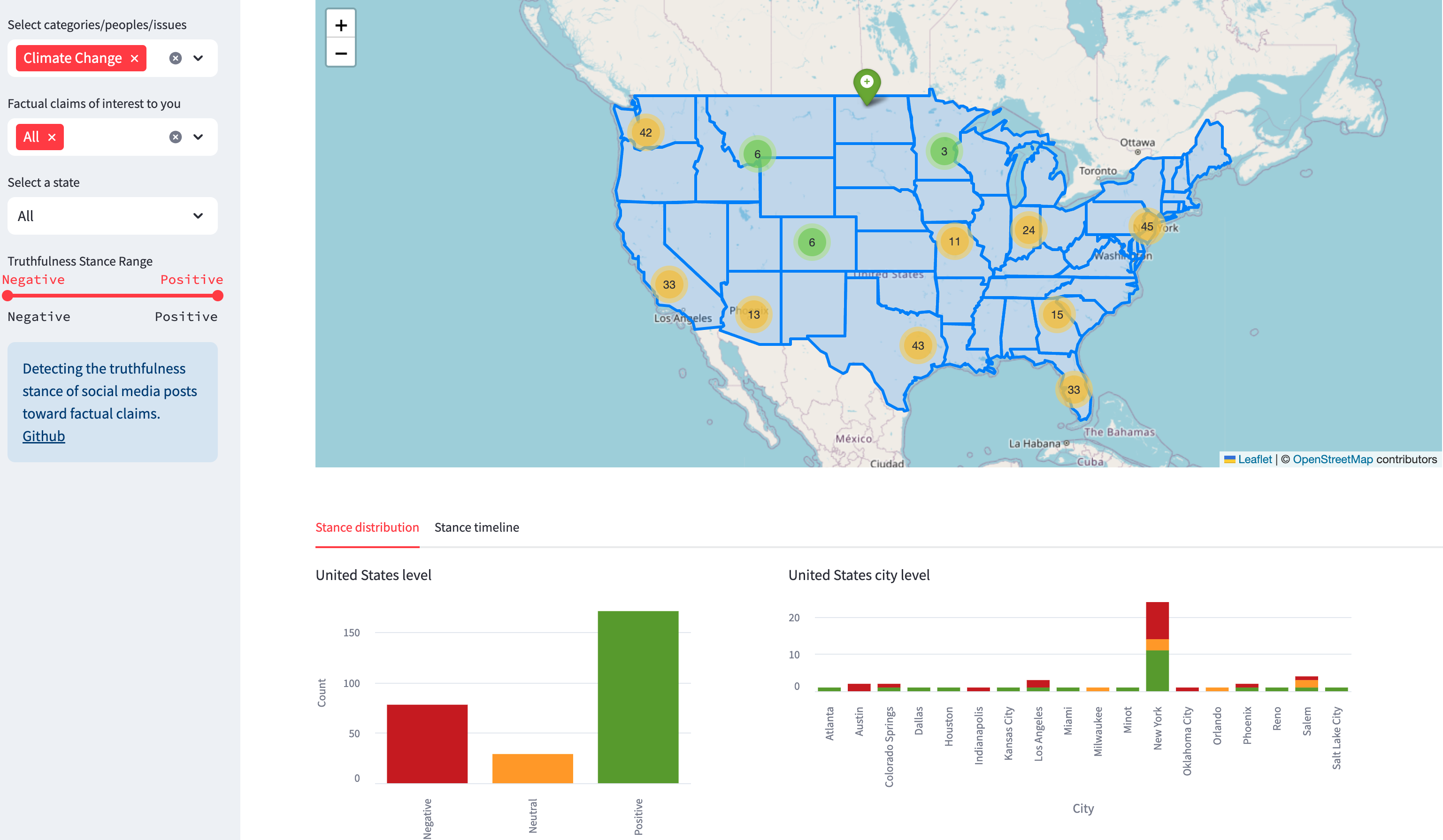}
		\caption{The user interface of \TRUSTMAP.}
		\label{fig:map}
    \end{center}
    \vspace{-6mm}
\end{figure*}

\section{Design of TrustMap}
\TRUSTMAP\ follows a three-stage pipeline in its design: (1) data ingestion, (2) truthfulness stance classification, and (3) data visualization.
In data ingestion, we collected factual claims from Politifact. We then formed claim-tweet pairs by querying X's API to retrieve claim-related tweets. \TRUSTMAP\ uses \RATSD\ to classify the truthfulness stance of each claim-tweet pair. Each claim-tweet pair is classified as either positive, negative, or neutral/no stance. In data visualization, these labeled claim-tweet pairs are grouped by their geographical locations, and stance distributions are rendered on an interactive map and through charts.

\subsection{Data Ingestion}

We collected factual claims from Politifact using an in-house fact-check collection tool.
% Note that only PolitiFact claims are included for user exploration within \TRUSTMAP\ due to X API rate limits and interface constraints. 
PolitiFact is chosen because it provides the most comprehensive coverage, with 25,694 claims spanning a wide range of topics.
For each claim, we retrieved tweets by constructing queries using a keyword-based retrieval strategy to extract key terms—typically nouns, verbs, and numbers—from the claim. 
Tweets were collected over a defined time window: one month before and up to one year after the publication date of the fact-check corresponding to each claim. Tweets shorter than 30 characters were excluded to reduce noise.

We used two strategies to collect geolocation information from the tweets. This information is key to rendering the claim-tweet pairs on \TRUSTMAP. The first strategy was to include location operators (e.g., latitude, longitude, and radius) in the query, e.g., ``[keywords] geocode:[latitude], [longitude], [radius]''. However, this strategy only returned limited data as X users rarely disclose their location in tweets. The second strategy involved collecting tweets and their metadata and then searching for geolocation information within the profiles of the X users who posted the tweets. The geolocation information was normalized using Geopy~\cite{geopy2023}, which converts unstructured location text into structured fields such as city, county, state, and country. In total, 136,040 claim-related tweets were collected for 2,216 unique claims, of which 24,262 tweets include usable geolocation data. The geographical distribution of the collected tweets can be found in Figure~\ref{fig:tweet_distribution} in the Appendix. 

\subsection{Truthfulness Stance Detection Model}
We applied \RATSD\ to classify the stance of each tweet toward its associated factual claim. \RATSD~ incorporates contextual knowledge to improve classification performance. This enhancement is critical because 1) both claims and tweets are standalone sentences that often lack sufficient context, making it difficult for a classification model to make an informed decision, and 2) tweets frequently contain acronyms, hashtags, and slangs, which pose challenges for the classification model to interpret accurately.

\RATSD\ enhances stance classification through two components---context retrieval and stance analysis generation. For each claim-tweet pair, the model first retrieves relevant contextual information using RAG.
It constructs contextual documents separately for each claim and each tweet, using claim-related and tweet-related data, respectively. This process involves selecting relevant documents, retrieving semantically aligned chunks using BGE embeddings~\cite{xiao2023c}, and prompting an LLM to summarize contextual information.

Next, \RATSD\ prompts an LLM to generate a stance analysis, which is a natural language explanation of the tweet's stance toward the claim. This analysis replaces the original tweet in the classifier’s input, allowing the model to learn from structured, context-rich representations instead of raw social media text.

Finally, \RATSD\ uses a fine-tuned LLM that encodes the input sequence consisting of claim, tweet, stance analysis, and contextual knowledge. The model outputs a stance category based on the probability distribution computed via a softmax layer. The model uses cross-entropy as loss function with L2 regularization, and parameters are optimized using the Adam optimizer~\cite{kingma2014adam}.

\subsection{Data Visualization}

The user interface (Figure~\ref{fig:map}) of \TRUSTMAP\ presents truthfulness stance results through an interactive map that supports filtering and exploration across different regions, topics, and claims. Built with Streamlit,\footnote{https://streamlit.io/} the interface consists of a control panel, an interactive map, and statistical charts for analyzing truthfulness stance classification results.

The control panel on the left allows users to select one or more topics. All claims within the chosen topics are selected. The users can further use a dropdown menu to choose specific factual claims belonging to the selected topics. 
%%%By default, the topic ``Climate Change'' is selected, and all claims within the chosen topic(s) are selected. 
Moreover, users can narrow the analysis to a certain region by selecting a state, either through the dropdown or by clicking directly on the map. Once a state is selected, the map will zoom in on that state, and the statistical charts will update to display the corresponding data. A stance slider over the spectrum of three stance categories---from Negative to Neutral/No Stance to Positive---provides an additional filter, enabling users to display claim-tweet pairs that fall within the chosen stance categories. 

The main area of the user interface contains a U.S. map and bar chart visualizations.
On the map, claim-tweet pairs are clustered based on their geographic coordinates.
% The map shows the total number of claim-tweet pairs in each region. 
As users zoom in, individual claim-tweet pairs appear as markers on their specific locations. Clicking on a marker opens a popup that displays the tweet, the associated claim, the claim’s verdict, and the detected truthfulness stance of the tweet toward the claim, as shown in Figure~\ref{fig:stance_example}.

Below the map, two bar charts summarize the stance distributions. The chart on the left displays the claim-tweet pair counts for each stance category at the national level or, if a state is selected, at the state level. The chart on the right presents stance distribution by city within the selected states or from across the entire United States if no state is selected. Users can also switch to the timeline tab to view a time series chart showing daily stance counts based on the selected claims. 
By integrating geographic and temporal views, the interface enables users to examine how X users' truthfulness stances vary across regions and change over time. 

%% file: sec-result.tex
\section{Results and Analysis}

\subsection{Performance Evaluation and Results}
\begin{table}[t]
  \small
  \centering
  \begin{tabular}{l@{\hspace{0.6em}}c@{\hspace{0.6em}}c@{\hspace{0.6em}}c@{\hspace{0.6em}}c}
    \toprule
    \textbf{Model} & $F_{\oplus}$ & $F_{\odot}$ & $F_{\ominus}$ & $F_{M}$ \\ \midrule
    BUT-FIT & 83.38 & 72.00 & 65.11 & 80.11 \\ 
    BLCU\_NLP & 85.37 & 71.43 & 63.29 & 73.36 \\ 
    BERTSCORE+NLI & 88.68 & 72.53 & 81.04 & 80.75 \\
    BART+NLI & 88.00 & 73.42 & 74.25 & 78.56 \\
    TESTED & 84.09 & 72.37 & 67.90 & 74.75 \\ \midrule
    RATSD\textsubscript{Zephyr} & 88.67 & 77.38 & 80.28 & 82.10 \\ 
    RATSD\textsubscript{GPT-3.5} & \textbf{93.27} & \textbf{80.24} & \textbf{87.90} & \textbf{87.13} \\
    \bottomrule
  \end{tabular}
  \caption{Performance comparison on the human-annotated dataset.}
  \label{tab:stance_score}
  \vspace{-2mm}
\end{table}

\begin{table}[t]
\centering
\resizebox{\columnwidth}{!}{
    \begin{tabular}{l|ccc|cc}
    \toprule
    \textbf{Stance} & \textbf{Truth} & \textbf{Mixed} & \textbf{Misinfo} & \textbf{Precision} & \textbf{F1} \\
    \midrule
    $\oplus$ & 6,754 & 5,094 & 64,643 & 9.0 & 15.6 \\
    $\odot$ & 1,398 & 1,350 & 9,677 & 10.9 & 11.7 \\
    $\ominus$ & 3,494 & 4,453 & 39,177 & 83.1 & 48.8 \\
    \midrule
    \textbf{Recall} & 58.0 & 12.4 & 34.6 & - & - \\
    \bottomrule
    \end{tabular}
}
\caption{Distribution of X users' truthfulness stances toward true, mixed, and false claims, along with the precision, recall, and F1-score for each stance category.}
\label{tab:stance_distribution}
% \vspace{-1mm}
\end{table}

\begin{table*}[t]
    \centering
    \small
    \begin{tabular}{l|cc|cc|c|c}
        \toprule
        \textbf{Topic} & \textbf{Truth-$\oplus$} & \textbf{Truth-$\ominus$} & \textbf{Misinfo-$\oplus$} & \textbf{Misinfo-$\ominus$} & \textbf{Accuracy} & \textbf{Macro F1} \\
        \midrule
        Public Health & 77.2\% (413) & 22.8\% (122) & 53.8\% (4,190) & 46.2\% (3,603) & 61.7 & 39.3 \\
        Elections & 40.0\% (8) & 60.0\% (12) & 58.9\% (4,582) & 41.1\% (3,201) & 40.6 & 29.3 \\
        Immigration & 54.1\% (53) & 45.9\% (45) & 60.6\% (3,023) & 39.4\% (1,969) & 46.8 & 29.8 \\
        Economy & 69.5\% (534) & 30.5\% (234) & 57.4\% (1,174) & 42.6\% (872) & 56.1 & 49.2 \\
        Abortion & \color[HTML]{036400}\textbf{79.9\% (446)} & \color[HTML]{036400}\textbf{20.1\% (112)} & 44.3\% (759) & 55.7\% (953) & \color[HTML]{036400}\textbf{67.8} & \color[HTML]{036400}\textbf{59.6} \\
        Education & 57.0\% (231) & 43.0\% (174) & 50.7\% (849) & 49.3\% (827) & 53.2 & 46.4 \\
        % Federal Budget & 77.9\% (345) & 22.1\% (98) & 54.8\% (945) & 45.2\% (780) & 61.5 & 49.9 \\
        Crime & 78.2\% (223) & 21.8\% (62) & \color[HTML]{680100}\textbf{79.5\% (1,030)} & \color[HTML]{680100}\textbf{20.5\% (265)} & 49.4 & 30.8 \\
        Environment & \color[HTML]{680100}\textbf{32.1\% (131)} & \color[HTML]{680100}\textbf{67.9\% (277)} & \color[HTML]{036400}\textbf{41.0\% (661)} & \color[HTML]{036400}\textbf{59.0\% (953)} & \color[HTML]{680100}\textbf{30.0} & \color[HTML]{680100}\textbf{28.6} \\
        \midrule
        All & 66.0\% (6,754) & 34.0\% (3,494) & 62.3\% (64,643) & 37.7\% (39,177) & 51.8 & 35.0 \\
        \bottomrule
    \end{tabular}
    \caption{Truthfulness stance distribution towards \textbf{Truth} and \textbf{Misinfo}rmation across PolitiFact's topics. Truth-$\oplus$ and Truth-$\ominus$ denote positive and negative stances towards \textbf{Truth}, respectively. Misinfo-$\oplus$ and Misinfo-$\ominus$ denote positive and negative stances towards \textbf{Misinfo}rmation, respectively. Note that “All” includes every topic, not just the selected eight, so the total count for the eight topics does not equal the count for ``All.''}
    \label{tab:topic_stance}
    \vspace{-4mm}
\end{table*}

The \RATSD~model was trained on the dataset from \citet{ratsd-naacl25}, which consists of 2,220 labeled claim-tweet pairs. The dataset includes 1,262 positive stance examples (stance category denoted as $\oplus$), 451 neutral/no stance examples ($\odot$), and 507 negative stance examples ($\ominus$). 
We evaluated the performance of \RATSD\ by comparing it to several state-of-the-art stance detection models, including fine-tuned LMs such as pre-trained models (e.g., BUT-FIT~\cite{fajcik2019but}), generative pre-trained models (e.g., BLCU\_NLP~\cite{yang2019blcu_nlp}), and domain-adaptive pre-trained models (e.g., BERTSCORE+NLI~\cite{hossain-etal-2020-covidlies}, BART+NLI~\cite{reddy2021newsclaims} and TESTED~\cite{Arakelyan2023TopicGuidedSF}). In \RATSD, we utilize two fine-tuned LLMs as alternative choices: the open-source model Zephyr~\cite{tunstall2023zephyr} and the proprietary model GPT-3.5~\cite{brown2020language}. 
We used F1 scores for each class—denoted as $F_{\oplus}$, $F_{\odot}$, and $F_{\ominus}$—and the macro F1 score ($F_M$) as evaluation metrics.
The evaluation results in Table~\ref{tab:stance_score}, reproduced from~\cite{ratsd-naacl25}, show that \RATSD\ achieves strong performance across all stance categories. 

In \TRUSTMAP, the \RATSD\ model was applied to the collected claim-tweet pairs to predict stance categories. Given the model's strong accuracy in stance detection, we rely on the predicted stance categories to evaluate the accuracy of X users' judgments on claims. More specifically, given each claim-tweet pair, we compared PolitiFact's veracity verdict for the claim with the stance expressed in the tweet regarding the claim's veracity. PolitiFact assigns one of six verdicts to each claim: ``True,'' ``Mostly True,'' ``Half True,'' ``Mostly False,'' ``False,'' and ``Pants on Fire.'' 
For simplicity and clarity, as well as for avoiding overly small fragments of data, we mapped the first two to ``Truth,'' ``Half True'' to ``Mixed,'' and the latter three to ``Misinformation.'' A claim-tweet pair's stance is considered accurate if the stance aligns with the claim’s veracity---e.g., a positive stance toward truth, a negative stance toward misinformation, or a neutral/no stance toward a mixed-veracity claim. Following this rule, we computed the precision and F1 score of users' judgments for each stance category, as well as the recall for each verdict category of the claims (Table~\ref{tab:stance_distribution}). These scores illustrate the difficulty X users face in distinguishing between truth and misinformation. 

The results show a strong tendency for X users to believe claims are true, regardless of their actual veracity. These findings align with recent research~\cite{moravec2018fake, zhang-etal-2024-granular}. More specifically, nearly 57.0\% ($\frac{64,643}{64,643+9,677+39,177}$) of the misinformation is believed to be true by the users, and the recall for Misinfo is only  0.346. 
Furthermore, users show significant skepticism even toward true claims, as only 58.0\% of the tweets about true claims express a positive stance, and over a third believe they are false.
Mixed-veracity claims (those labeled ``Half True'') reveal another concerning pattern. Users rarely express neutral or uncertain stances toward such claims. Instead, they are often polarized, leaning toward either accepting them as true or rejecting them as false.

\subsection{Detailed Analysis}

\TRUSTMAP\ can be useful in identifying topics and regions where X users' perceptions toward true and false claims diverge from the veracity of claims themselves. Such insights can help guide more targeted fact-checking and public awareness efforts.
Therefore, we conducted analyses from both topic and geographical perspectives. 
For these analyses, we excluded a claim-tweet pair if the claim has a ``Mixed'' verdict and/or the tweet is classified as $\odot$. 
This enables us to derive clearer insights without accounting for the less frequent and more ambiguous stance and verdict categories.  
Moreover, we produced results separately for each of the top eight states in terms of number of claim-tweet pairs. Similarly, we produced results for top eight topics. Note that while topics such as ``National,'' ``Space,'' and ``Federal Budget'' are populous, they were not analyzed individually because they are either highly imbalanced (e.g., containing only misinformation with no truth) or do not truly constitute distinct topics.

\begin{table*}[ht]
    \centering
    \small
    % \resizebox{0.85\textwidth}{!}{
        \begin{tabular}{l|cc|cc|c|c}
        \toprule
        \textbf{Region} & \textbf{Truth-$\oplus$} & \textbf{Truth-$\ominus$} & \textbf{Misinfo-$\oplus$} & \textbf{Misinfo-$\ominus$} & \textbf{Accuracy} & \textbf{Macro F1} \\
        \midrule
        Washington & 74.1\% (177) & 25.9\% (62) & \color[HTML]{036400}\textbf{42.7\% (719)} & \color[HTML]{036400}\textbf{57.3\% (963)} & \color[HTML]{036400}\textbf{65.7} & \color[HTML]{036400}\textbf{51.2} \\
        Florida & 69.4\% (120) & 30.6\% (53) & \color[HTML]{680100}\textbf{66.6\% (1,147)} & \color[HTML]{680100}\textbf{33.4\% (576)} & \color[HTML]{680100}\textbf{51.4} & \color[HTML]{680100}\textbf{32.8} \\
        Texas & 70.3\% (204) & 29.7\% (86) & 62.7\% (1,006) & 37.3\% (599) & 53.8 & 39.8 \\
        New York & 75.4\% (138) & 24.6\% (45) & 54.9\% (850) & 45.1\% (698) & 60.3 & 42.3 \\
        California & 67.1\% (94) & 32.9\% (46) & 62.7\% (827) & 37.3\% (491) & 52.2 & 35.3 \\
        Arizona & \color[HTML]{036400}\textbf{78.0\% (46)} & \color[HTML]{036400}\textbf{22.0\% (13)} & 63.3\% (321) & 36.7\% (186) & 57.3 & 37.1 \\
        Colorado & \color[HTML]{680100}\textbf{66.7\% (24)} & \color[HTML]{680100}\textbf{33.3\% (12)} & 58.1\% (299) & 41.9\% (216) & 54.3 & 35.8 \\
        Virginia & 73.2\% (60) & 26.8\% (22) & 62.3\% (288) & 37.7\% (173) & 55.3 & 40.3 \\
        \midrule
        United States & 70.0\% (1,444) & 30.0\% (619) & 59.9\% (10,299) & 40.1\% (6,886) & 55.0 & 38.3 \\
        All & 65.9\% (6,754) & 34.1\% (3,494) & 62.3\% (64,643) & 37.7\% (39,177) & 51.8 & 35.0 \\
        \bottomrule
        \end{tabular}
    % }
    \caption{Truthfulness stance distribution toward \textbf{Truth} and \textbf{Misinfo}rmation across U.S. states.}
    \label{tab:state_stance}
\end{table*}

\begin{table*}
\centering
\small
% \resizebox{\textwidth}{!}{
    \begin{tabular}{l|cc|cc|c|c}
    \toprule
    \textbf{Political Leaning} &  \textbf{Truth-$\oplus$} &  \textbf{Truth-$\ominus$} &  \textbf{Misinfo-$\oplus$} &  \textbf{Misinfo-$\ominus$} &  \textbf{Accuracy} &  \textbf{Macro F1} \\
    \midrule
       Red States & \color[HTML]{680100}\textbf{67.9\% (527)} & \color[HTML]{680100}\textbf{32.1\% (249)} & \color[HTML]{680100}\textbf{63.7\% (4,344)} & \color[HTML]{680100}\textbf{36.3\% (2,476)} & \color[HTML]{680100}\textbf{52.1} & \color[HTML]{680100}\textbf{35.3} \\
       Blue States & 70.8\% (659) & 29.2\% (272) & \color[HTML]{036400}\textbf{55.6\% (4,259)} & \color[HTML]{036400}\textbf{44.4\% (3,402)} & \color[HTML]{036400}\textbf{57.6} & \color[HTML]{036400}\textbf{41.3} \\
       Swing States & \color[HTML]{036400}\textbf{73.4\% (257)} & \color[HTML]{036400}\textbf{26.6\% (93)} & 62.5\% (1,513) & 37.5\% (908) & 55.5 & 38.7 \\
    \bottomrule
    \end{tabular}
% }
\caption{Truthfulness stance distribution by political leaning.}
\label{tab:leaning_summary}
\vspace{-4mm}
\end{table*}

\paragraph{Topic-level analysis.}
Table~\ref{tab:topic_stance} shows how X users perceive factual claims across different topics. For each topic, we report the proportions of positive and negative stances toward both true and false claims, as well as the accuracy and macro F1 score to assess alignment between user stance and claim veracity.
The results show that in the topic of ``\textit{Crime},'' users express high levels of belief in both true and false claims, with 78.2\% for Truth-$\oplus$ and even the higher 79.5\% for Misinfo-$\oplus$. Conversely, the topic ``\textit{Environment}'' shows the highest rate of skepticism, with 67.9\% and 59.0\% of tweets rejecting truth and misinformation, respectively.

Across topics, accuracy and macro F1 scores vary, but most fall below 0.5. ``\textit{Abortion}'' has the highest scores (Accuracy: 67.8, F1: 59.6), indicating relatively better user judgment in this domain. In contrast, ``\textit{Environment}'' and ``\textit{Elections}'' exhibit the lowest F1 scores (28.6 and 29.3), suggesting limited user ability to differentiate between true and false claims related to these topics.

\paragraph{Geographical-level analysis.}
Similarly, we analyzed X users' stance by geographical location. As shown in Table~\ref{tab:state_stance}, both tweet count and percentage for Misinfo-$\oplus$ reach the highest (66.6\%) in Florida, meaning that misinformation is widely spread in this region. In addition, Florida X users exhibit the lowest accuracy (51.4) and macro F1 score (32.8).
In contrast, Washington stands out with the highest accuracy (65.7) and macro F1 score (51.2), as well as the lowest Misinfo-$\oplus$ (42.7\%), indicating in Washington relatively more X users push back against misinformation.
Even in higher-performing states like Washington, however, F1 scores remain modest. This suggests that, while some regional variation exists, users overall struggle to discern the truthfulness of claims. This observation is reinforced by the national-level (``United States'') accuracy at 55.0 and macro F1 at just 38.3.
When comparing ``United States'' with ``All'', which includes claim-tweet pairs from within the U.S., outside the U.S., and those without geolocation, we find that users in the United States perform slightly better at distinguishing between true and false claims, as reflected in their higher accuracy and macro F1.

We also uncovered an interesting finding regarding political leaning. Similar to Table~\ref{tab:state_stance}, Table~\ref{tab:leaning_summary} presents the distributions of truthfulness stances across ``red'' (i.e., leaning toward Republicans), ``blue'' (i.e., leaning toward Democrats), and ``swing'' (i.e., no leaning) states. The classification of red, blue and swing states is based on information from Wikipedia.\footnote{\url{https://en.wikipedia.org/wiki/Red_states_and_blue_states}} The table shows X users in blue states are better at distinguishing between true and false information compared to those in red states. This result is consistent with the findings from prior research in political science and psychology~\cite{garrett2021conservatives,dobbs2023democrats,spampatti2024conservatives,zhu2025political}.

% Our analyses highlight a consistent pattern: X users frequently believe claims are true regardless of their actual veracity. These findings align with recent research~\cite{moravec2018fake, zhang-etal-2024-granular} showing that individuals often fail to distinguish between real and fake content.
% In addition, in both topic- and state-level evaluations, we found that X users' judgments of factual claims vary across topics and regions. 

%% file: sec-conclusion.tex
\section{Conclusion}
\TRUSTMAP\ is an interactive visualization tool that helps analyze how X users across different regions respond to factual claims. By combining truthfulness stance detection with geospatial and topical breakdowns, the system provides a detailed view of public reactions to both true and false information. The results reveal clear variations in how misinformation is received across topics and locations. This work demonstrates the value of pairing automated truthfulness stance classification with visual analytics to support monitoring and response efforts in the fact-checking ecosystem.

\section*{Limitations}
While \TRUSTMAP\ offers a novel interface to visualize truthfulness stance toward factual claims across the U.S., it has several limitations. 
The reliance on X's  API introduces data access constraints. Due to API rate limits and restrictions on historical tweet access, the dataset may not comprehensively capture all relevant tweets for every claim. Additionally, many tweets lack precise geolocation data, which will lead to potential sampling bias. The geolocation inference method based on user profiles introduces further noise, as profile locations are user-defined and may not reflect actual user locations.
Additionally, while \TRUSTMAP\ presents static snapshots of public discourse, it does not support real-time data updates, which would be necessary for early misinformation detection and timely interventions.

\section*{Ethics Statement}
This work raises ethical considerations related to user privacy and content sensitivity. Although all analyzed data is publicly available from X, user tweets may contain sensitive or personally identifiable content. We take steps to minimize exposure of private information by displaying only essential tweet content and omitting usernames, user profiles, and direct user identifiers.
Additionally, stance classification results are generated automatically using LLMs. These predictions may not always reflect the intent of the original X users and could misrepresent X users' views. Users of the system should interpret the results as aggregated patterns, not definitive assessments of individual posts.
Finally, while \TRUSTMAP\ is intended to support fact-checking and public awareness efforts, there is a risk of misuse if the system is interpreted as providing a definitive judgment on the truthfulness of claims.
Our tool is designed to support transparency and encourage informed evaluation, not to police online discourse.

%% file: sec-appendix.tex
\appendix

% \begin{table*}[t]
%     \small
%     \centering
%     \resizebox{\textwidth}{!}{
%         \begin{tabular}{lccccccc}
%             \toprule
%             \textbf{Fact-check Components} & AFP Fact Check & AP Fact Check & FactCheck.org & FullFact & Metafact & PolitiFact & Snopes \\ 
%             \midrule
%             Claims & 0\textsuperscript{*} & 3,127 & 0\textsuperscript{*} & 3,995 & 3,428 & 25,694 & 18,198 \\ 
%             Review Summaries & 8,118 & 3,122 & 3,797 & 3,995 & 0\textsuperscript{*} & 25,679 & 2,651  \\ 
%             Reviews & 8,223 & 3,290 & 3,797 & 3,995 & 3,428 & 25,686 & 19,285 \\ 
%             Verdicts & 0\textsuperscript{*} & 3,127 & 0\textsuperscript{*} & 0\textsuperscript{*} & 3,428 & 25,694 & 14,009 \\ 
%             \bottomrule
%         \end{tabular}
%     }
%     \caption{Counts of claims, review summaries, full reviews, and verdicts in the fact-check collection. Asterisks (*) indicate data types not provided by the corresponding source.}
%     \label{tab:factcheck_dataset}
% \end{table*}
\todo[color=pink]{In Table~\ref{tab:factcheck_dataset}, instead of o*, why not just use N/A? Replace it in caption too. | Table is commented out.}

\section{Appendix}
\label{sec:appendix}

% \subsection{Details of Claim Data}
% Table~\ref{tab:factcheck_dataset} summarizes the contents of our claim data. For each source, we report the number of claims, review summaries, full review articles, and verdicts collected.

\subsection{Geographic Distribution of Collected Tweets}
\begin{figure}[ht]
    \begin{center}
    	\includegraphics[width=1.0\columnwidth]{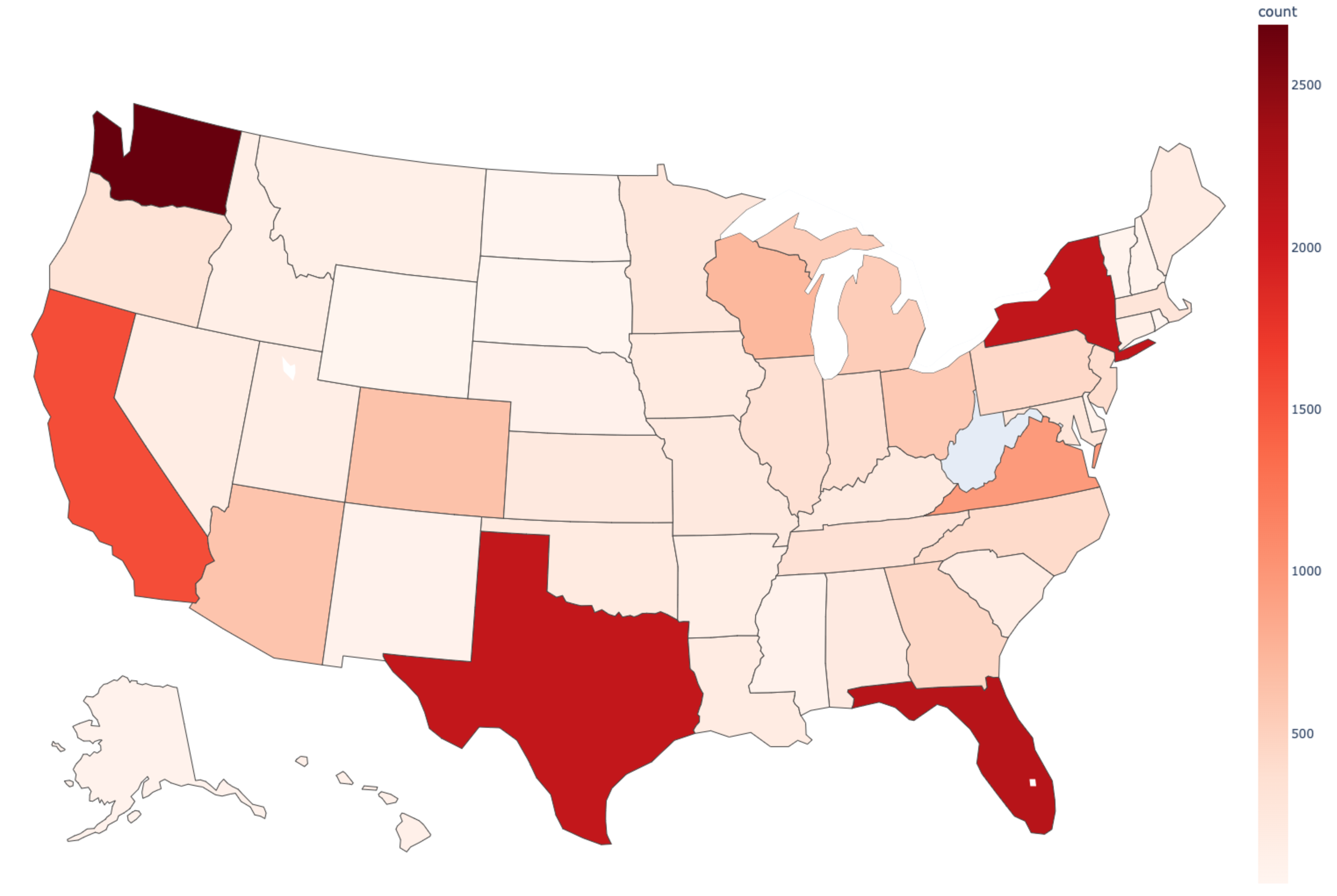}
		\caption{Geographic distribution of collected tweets across U.S. states.}
		\label{fig:tweet_distribution}
    \end{center}
\end{figure}

Figure~\ref{fig:tweet_distribution} shows a heatmap illustrating the geographic distribution of collected tweets across U.S. states. Darker shades represent higher tweet volumes. States such as Texas, Florida, California, and New York show the highest levels of tweet activity. In contrast, several central and mountain states, including the Dakotas, Wyoming, and Montana, show relatively sparse activity.\todo{Remove the last two sentences about states.}

